\documentclass[runningheads]{llncs}
\usepackage[T1]{fontenc}
\usepackage{makeidx}  %
\usepackage{microtype}
\usepackage{tabularx}
\usepackage[table]{xcolor}
\usepackage{booktabs}
\usepackage{todonotes}
\usepackage{amsmath}
\usepackage{amssymb}
\usepackage{pifont}
\usepackage{cite}
\usepackage{color}
\usepackage[colorlinks=true,linkcolor=blue,urlcolor=blue,citecolor=blue]{hyperref}
\usepackage{cleveref}
\usepackage{paralist}
\usepackage[section]{placeins}

\newcommand{\mypar}[1]{\smallskip\noindent\textbf{#1.}}

\begin{document}
\mainmatter              %
\title{Towards Nudging in BPM: A Human-Centric Approach for Sustainable Business Processes}
\titlerunning{Nudging for Sustainable Business Processes}  %
\author{Cielo González Moyano\inst{1} \and Finn Klessascheck\inst{2,3} \and Saimir Bala\inst{1} \and Stephan A. Fahrenkrog-Petersen\inst{1,3} \and Jan Mendling\inst{1,3}}
\authorrunning{C. González Moyano et al.}   %
\institute{Humboldt-Universität zu Berlin, Berlin, Germany\\
\email{firstname.lastname@hu-berlin.de}\\ 
\and
Technical University of Munich, School of CIT, Heilbronn, Germany\\\email{firstname.lastname@tum.de} \and
Weizenbaum Institute, Berlin, Germany
}

\maketitle              %

\begin{abstract}        %
Business Process Management (BPM) is mostly centered around finding technical solutions. Nudging is an approach from psychology and behavioral economics to guide people's behavior. 
In this paper, we show how nudging can be integrated into the different phases of the BPM lifecycle. Further, we outline how nudging can be an alternative strategy for more sustainable business processes.
We show how the integration of nudging offers significant opportunities for process mining and business process management in general to be more human-centric. 
We also discuss challenges that come with the adoption of nudging.
\end{abstract}

\section{Introduction}
Business Process Management (BPM) aims at governing business processes throughout their lifecycle, with a continuous focus on their improvement. Business processes describe work undertaken for achieving specific goals for the organization, such as delivering a product to the customer. To this end, BPM develops novel techniques for the analysis and optimization of all aspects related to this work. An established category of process analysis techniques is process mining~\cite{van2012process}, which has the goal of providing data-driven insights about the process.

Research on process mining and other BPM techniques focuses on event data and the development of techniques to extract insights and recommendations from this data.
As a result, a number of highly technical solutions to a various BPM problems have been developed.
In many scenarios, considering human aspects can bring additional opportunities for process improvements. 
This is observed, among others, in the space of prescriptive process monitoring~\cite{DBLP:conf/bpm/DeesLAR19,DBLP:journals/kais/Fahrenkrog-Petersen22}, where human actions shall be triggered based on process predictions. 
In other words, while the techniques developed in BPM may provide important actionable insights into the process, they become futile if not put into practice by stakeholders. 
Furthermore, by focusing on optimization, BPM endeavors often overlook their environmental impact\cite{DBLP:journals/ajis/GhoseHHL09,seidelGreenBusinessProcess2012}.

To address this issue, we propose the integration of nudges~\cite{thaler2021nudge,DBLP:journals/bise/WeinmannSB16} into BPM.
Take for example the process of requesting, conducting, and reimbursing business trips, which is important for organizations so that they can correctly handle and recompense travel costs~\cite{pufahl2020performance}. Notably, business travel, which has been linked with facilitating innovation and technology transfer~\cite{hovhannisyanInternationalBusinessTravel2015}, is predicted to account for 1 to 4 percent of global greenhouse gas emissions in 2050~\cite{mccainBusinessTravelGHG2021}. Through nudging, employees requesting and reimbursing their business travels can be guided towards certain behavior that may contribute towards reducing the emission of business travel, and thus, the environmental impact. For example, they might be nudged to 1) conduct non-essential or domestic business trips via rail instead of plane (i.e., a mode of travel that causes fewer emissions), 2) voluntarily reimburse CO\textsubscript{2} emissions from their flights~\cite{DBLP:conf/ecis/SzekelyWB16}, 3) forego some business trips entirely in favor of e.g. video-conferencing, or 4) book flights that may be unfavorable for their travel experience, but cause fewer emissions, and 5) request reimbursement digitally instead of via a printed form.

\begin{figure}
\vspace{-1em}
    \centering
    \includegraphics[width=0.8\linewidth]{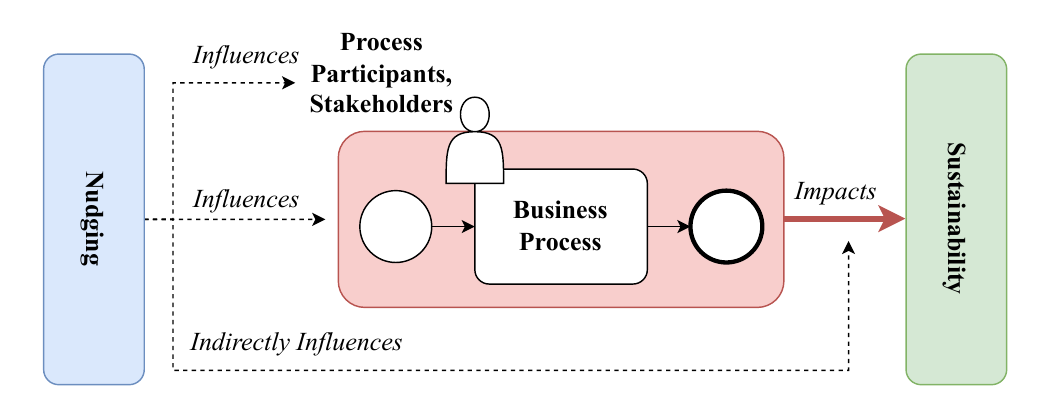}
    \caption{Conceptual overview of Nudging for Sustainable Business Processes}
    \label{fig:nudge4s}
\end{figure}

In this work, we will show how nudging can be integrated into the BPM lifecycle to guide an organization, its stakeholders and process participants towards better and environmentally aware business processes~\cite{klessascheck2024unlockingsustainabilitycompliancecharacterizing}. Figure~\ref{fig:nudge4s} illustrates the idea of nudging for sustainable business processes and the main concepts we draw upon: By influencing process participants and stakeholders to take sustainability issues into account, nudging influences an organization's business processes to be more environmentally aware. Further, by integrating sustainability concerns into business processes and their management, nudging ultimately influences the way an organization and its business processes impact sustainability.

The remainder of the paper is structured as follows. In \Cref{sec:background}, we provide a background on nudging and the different types of nudges. 
In \Cref{sec:nudge}, we show how the nudges can be integrated into the different phases in the BPM lifecycle. \Cref{sec:nudge} we highlight potential research directions for process mining researchers in the area of digital nudges. In \Cref{sec:discussion}, we discuss the implications of integrating nudging into BPM. In \Cref{sec:conclusion} we draw the conclusions.

\section{Background}
\label{sec:background}

Only a few works explore nudging in BPM~\cite{DBLP:journals/ajis/GhoseHHL09,DBLP:journals/bpmj/BammertKRW20}. 
In the following, \Cref{subsec:what-is-nudging} introduces nudging, \Cref{subsec:why-nudging} discusses its relevance on decision-making in BPM, and \Cref{subsec:nudging-for-green} highlights its potentials for Green BPM.

\subsection{What Is Nudging?}\label{subsec:what-is-nudging}
Nudging is a term coined in behavioral economics and social psychology where people are subtly guided toward desired behaviors or decisions without restricting their choices or significantly changing the incentives~\cite{thaler2021nudge, sunstein2015ethics}. A nudge increases the probability that an individual will make a specific choice or act in a particular way by modifying the environment to trigger an automatic cognitive response that favors a desired outcome~\cite{thaler2021nudge}. To be considered as a nudge, the intervention must be simple and allow people to make their decisions~\cite{sunstein2015ethics}. Table~\ref{tab:categoriesNudges} presents a categorization of the nudges based on some of the strategies that are employed to influence behavior. We briefly describe these the various types of nudge below.

\begin{table}
\caption{Types of nudges to guide user behavior}
\vspace{-0.2cm}
\label{tab:categoriesNudges}
\centering
\begin{tabular}{ll}
    \toprule
    \textbf{Category} & \textbf{Description} \\
    \midrule
   \textit{Default effect}  & Automatically select a particular option.\\
   \textit{Social norms} & Influence individual behavior according to the group behavior.\\
    \textit{Framing effect} & Alter the interpretation of information by changing its presentation.\\
    \textit{Salience} & Capture attention by making certain choices more visually prominent.\\
    \textit{Priming} & Trigger certain associations in the mind.\\
    \textit{Simplification} & Reduce the complexity of a choice.\\
    \textit{Reminders} & Produce timely notifications that encourage to act.\\
    \textit{Feedback} & Provide information about behavior and its consequences.\\
    \bottomrule
\end{tabular}
\end{table}

A type of nudge is the \textit{default effect}, where a pre-selected option is chosen unless the individual actively chooses otherwise. Another type is the use of \textit{social norms}, making people conform to what they perceive others are doing. For instance, informing individuals that a high percentage of their neighbors are donating to a fund-raiser program can significantly increase the donations~\cite{cialdini2001science}. Similarly, \textit{framing effects}, where the way information is presented influences decision-making, is a commonly used type of nudge. For example, highlighting the benefits of taking a certain action (e.g., 90\% success rate) in comparison to another option (e.g., 10\% success rate) can influence the final decision~\cite{tversky1988rational}.

Nudges commonly work because they inform people or make their decisions easier in an environment designed with that intent~\cite{thaler2021nudge}. For instance, in \textit{salience} nudges, placing healthier foods at eye-level in a store can increase their selection without removing less healthy options~\cite{valenvcivc2024changing}. \textit{Priming}, on the other hand, uses unconscious associations. Such associations have led participants in a diet study to change their eating behavior by being subtly exposed to diet-related words~\cite{papies2012goal}.
\textit{Simplification} reduces the complexity of decisions, making it easier for individuals to make beneficial choices. An example is the simplification of health insurance plan options, where clear, concise information helps people choose plans that best meet their needs without feeling overwhelmed by complexity~\cite{handel2013adverse}. In a similar health context, an example for \textit{reminders} is the notifications sent to patients to remind them to take medication, which has been shown to improve health outcomes~\cite{pop2011mobile}. As sending reminders brings advantages, providing \textit{feedback} is also a nudging strategy that can positively influence behavior. For instance, providing households with feedback on their energy consumption compared to their neighbors has led to a significant reduction in consumption~\cite{allcott2011social}.

All the nudges depend on creating an environment that addresses the behavior simply and economically without limiting the choices~\cite{thaler2021nudge, sunstein2015ethics}. To create this environment, nudging relies on an understanding of human behavior to design interventions that are both effective and ethical~\cite{bicchieri2022nudging}. In this context, it cannot be considered a nudge if the intervention imposes significant restrictions on choices or introduces substantial economic incentives. For example, a mandatory policy that requires people to participate in a program is not a nudge given that it removes individual choice~\cite{sunstein2015ethics}. Similarly, offering large financial incentives to adopt a specific behavior, such as substantial subsidies for purchasing electric cars, does not qualify as a nudge. Although these alternatives are also effective, they align with traditional economic strategies rather than with nudging~\cite{thaler2015misbehaving}.

\subsection{Why Is Nudging Relevant?}\label{subsec:why-nudging}
In their work, Thaler and Sunstein~\cite{thaler2021nudge} argue that the necessity of nudging arises from fundamental aspects of human nature. Two systems characterize human thinking: the \textit{reflective system} and the \textit{automatic system}~\cite{kahneman2011thinking}. The first one, the \textit{reflective system}, is deliberate and conscious. It requires individuals to process information carefully and consider the consequences of their actions. This system is typically associated with a slow and rigorous decision-making process. 

On the other hand, the \textit{automatic system} is quick and intuitive. It relies on heuristics to make decisions with minimal cognitive effort~\cite{kahneman2011thinking}. This system is typically used to make decisions that do not require exhaustive processing. However, humans do not always make complex decisions based on rationalization~\cite{gigerenzer2011heuristic}. When situations are highly complex, uncertain, or overwhelming, and there is pressure to make a decision, humans tend to rely on the \textit{automatic system}~\cite{tversky1974judgment, thaler2021nudge, kahneman2011thinking}. This results in faster decisions, but it can also lead to inadequate outcomes by relying on heuristics and biases or simply giving in to pressure.

Nudging strategies can effectively target these inadequate outcomes and support automatic decision-making by subtly guiding behavior through small changes in the environment or choice of architecture~\cite{thaler2021nudge,kahneman2011thinking}. This makes nudging particularly powerful, aligning with how people naturally think and make decisions. Using nudging, it is possible to achieve various BPM goals~\cite{DBLP:journals/bpmj/BammertKRW20}, such as improving productivity with cost-effective approaches~\cite{thaler2021nudge}, promoting healthy and sustainable behaviors~\cite{pop2011mobile, allcott2011social}, and executing scale interventions with technology~\cite{weinmann2016digital}.

\subsection{What Can Nudging Do for Green BPM?}\label{subsec:nudging-for-green}
Considering the environmental impact of businesses and their processes has become increasingly important, especially in light of climate change~\cite{roohygoharEnvironmentalSustainabilityGreen2020}. As a response, the field of \textit{Green BPM} has emerged, which explicitly aims to incorporate the issue of environmental sustainability into the traditional notion of BPM~\cite{seidelGreenBusinessProcess2012} in order to increase the environmental performance of organizations~\cite{roohygoharEnvironmentalSustainabilityGreen2020}.

Concretely, Green BPM is concerned with the modeling, deployment, optimization, and management of business processes in a way that particularly takes consequences on the environment into account. Further, organizational aspects beyond technical capabilities, such as culture (here, related to questions such as how ``green'' values can be communicated or individual stances of an organization's members can be influenced to favor less environmentally impactful behavior~\cite{seidelGreenBusinessProcess2012}), play an important role for a successful adoption of Green BPM~\cite{couckuytSystematicReviewGreen2019}.
However, existing research has mostly  focused on technical aspects, such as modeling capabilities, with little focus on organizational culture~\cite{couckuytGreenBPMBusinessOriented2019,fritschPathwaysGreenerPastures2022}.

As a consequence, we see significant potential in approaches such as nudging, in order to engage with BPM in general, and Green BPM in particular, in a way that goes beyond technical concerns and addresses cultural aspects. Notably, this would contribute to paving the way for further research on how attitudes and behaviors related to sustainable business processes can be shaped.

\section{Nudging and the BPM Lifecycle} \label{sec:nudge}
The BPM lifecycle consists of six phases: identification, discovery, analysis, redesign, implementation, and monitoring~\cite{DBLP:books/sp/DumasRMR18}. 
This section explores how nudging can be incorporated into each phase, providing a comprehensive framework for a human-centric and sustainable BPM approach.

\subsection{Identification}
As a first step, a process needs to be \emph{identified}. The goal of this phase is to define what a specific process is. 
To this end, existing knowledge of what work is being carried out in the organization has to be rationalized. 
This requires managers to think about the status quo of how work is carried out in the organization. 
Before any BPM initiative occurs, knowledge about the processes may be fully unstructured and not explicit. 
Nudging can come into play to aid BPM initiatives' success by guiding an organization to learn and adopt BPM concepts and methods.\\

\noindent\textbf{Helping Process Thinking.} A starting point for process identification is \emph{process thinking}. Typical process thinkers are able to identify the various ingredients of processes, such as actors, customers, value delivered to customers, and its possible outcomes. 
To change an organization's mindset toward process thinking, challenges must be faced. One big challenge is the \emph{status quo} bias. 
This represents a cognitive challenge and can be addressed by designing specific learning nudges~\cite{damgaard2018nudging}. 

In more practical terms, processes are often pre-defined and identified within the functional silos of an organization. Nudging is used to break up this silos and promote the end-to-end vision of the process. To this end, different nudging strategies can be applied. For instance, \emph{framing} the benefits of integrated processes in terms of improved efficiency and satisfaction of process participants can motivate stakeholders to adopt a broader view. \emph{Priming} can be used to subtly influence stakeholders to think beyond their immediate functional areas. For instance, workshops and training sessions can start with examples and case studies that emphasize the importance and benefits of cross-functional processes. \emph{Simplifying} the language and tools for describing and modeling processes can make it easier for users to create a shared understanding of the functions. 

Once there is a holistic view of the process, nudging can help ensure that sustainability considerations are included. For instance, \emph{framing} the business problem in terms of environmental impact can nudge stakeholders to consider sustainable practices when identifying relevant processes. Additionally, \emph{default options} can prioritize the processes that have the potential for the highest environmental benefits. This can guide the selection of processes for the next phases of the BPM lifecycle.

\noindent\noindent\textbf{Avoiding Scoping Mistakes.} Scoping the process is key to the identification phase. Not any arbitrary chunk of work is considered a business process.
To avoid mistakes in this phase, the BPM team takes a structured approach. Questions posed here are the following: Is it a process at all? Can the process be controlled? Is the process important enough to manage? Is the scope of the process not too big or not too small? Challenges are associated with all these questions. 

One big problem in this phase is that the scope of the process is inappropriate, resulting in processes that are not in line with the organization's values. Nudges such as \emph{framing}, \emph{feedback}, and \emph{social norms} are all useful in this phase. 
In the case of \textit{framing}, presenting the criteria for process identification within the context of the goals of the organization, can help stakeholders to align their scope definitions. For example, framing the discussion around how the process contributes to sustainability goals can help ensure that the scoped processes are environmentally aligned. 
Implementing \textit{feedback} loops where the initial scoping decisions are assessed can help in refining the scope over time. Providing feedback on the effectiveness and alignment of scoped processes with sustainability organizational values can nudge the BPM team to make necessary adjustments. With a similar approach, highlighting examples of well-scoped processes from within the organization or industry can serve as \textit{social norms} that nudge the BPM team towards best practices. Sharing success stories and case studies where proper scoping led to significant improvements can create normative pressure to follow similar approaches.

\subsection{Discovery}
In the process discovery phase, the current state of relevant processes is documented, typically in the form of as-is process models. This can be done both through manual work, and automatically through process mining.\\ 
\noindent\textbf{Mitigating Intrusiveness in Observation-based Process Discovery.}
One method used for process discovery is the observation of the process workers. This method is often challenging since process workers find it intrusive and might behave differently when under observation. 
This issue can be mitigated by applying nudging techniques. First, an opt-out protocol can be used, where process workers by \emph{default} participate in the observational study but can decide against their participation.

Furthermore, process workers might try to hide behavior that is considered unwanted. By explicitly encouraging them to show their workaround~\cite{DBLP:conf/bpm/BeerepootWR19a} a \emph{social norm} could be introduced that ensures a feeling of safety necessary to understand how they act within the business process. Additionally, before the observation begins, \emph{priming} workers with positive examples of how their input can lead to beneficial changes can help reducing anxiety and promoting honest behavior. This can be achieved via briefings or informational material.

\noindent\textbf{Preventing Manipulation of the Event Logs.}
Automated techniques like process mining offer a less intrusive alternative to manual observation. However, process mining depends on event logs, which in turn depend on the interaction of the users with the system. Users might manipulate these logs or the interactions with the system to present a more favorable view of their actions \cite{baier2018matching, vsinik2023peek}. In this context, providing users with regular \emph{feedback} on how their accurate logging has contributed to process improvements can reinforce positive behavior and discourage manipulation. 

\subsection{Analysis}
In the \textit{analysis} phase a weaknesses of the processes are identified and reported, providing information about its alignment with the business goals.

\noindent\textbf{Defining Goal-settings in Business Processes.}
The way goals are structured in a business can serve as a nudge. As an example, the aim could be that all processes produce zero waste, this would set a \emph{primer} on what analytical questions might be asked in the analysis phase. Furthermore, certain goals can be the \emph{default}, but it could be allowed to deviate from them, if they do not impact the goal of the business process, i.e., a zero-emissions goal might not be compatible to a process that requires employees to travel overseas within a set time frame.

\noindent\textbf{Identifying Issues with the As-Is Process.}
Providing immediate \emph{feedback} on the environmental performance of processes can motivate stakeholders to identify and prioritize issues related to sustainability. Comparative \emph{feedback}, such as benchmarking against industry standards or best practices, can further drive improvement efforts.

\noindent\textbf{Identifying Potential Use Cases for Nudges for Green BPM.}
Nudges can also be integrated into the business process, and the analysis phase can be used to identify the respective entry points. As an example, setting the \emph{default} option in a process to sign a document electronically instead of requiring a signature on paper, or to book a means of transport with generally lower emissions for a business trip. The main task in the analysis point would be to identify areas where customers prefer non-sustainable options, and nudges could be used to encourage more sustainable behavior.

\subsection{Redesign}
The purpose of the redesign phase is to produce a newly designed process model, that overcomes the issues identified in the analysis phase. To this end, we identify two situations where nudging can help. 

\noindent\textbf{Tackling the Status Quo.} Although there is a newly devised process model, resulting from the redesign, status quo bias within the organization can hinder its adoption by employees. Nudges can be designed to tackle this bias. For example, \emph{default} settings can make the steps of the to-be process the first option. Additionally, \emph{framing} can be used to highlight how the adoption of the to-be process improves certain issues or key performance indicators (KPIs). This can help shift the perception of the new process from a disruptive change to a beneficial enhancement.

\noindent\textbf{Encouraging Sustainable Redesign Options.}
Nudging can promote the consideration of sustainable redesign options. Presenting \emph{default} redesign templates that incorporate green practices can guide stakeholders toward more sustainable solutions.

\noindent\textbf{Validating Redesign Choices.}
Leveraging \emph{social norms}, redesign workshops can include examples of successful sustainable redesigns from other organizations. This can create a social expectation to adopt similar practices, thus nudging stakeholders toward environmentally friendly solutions

\noindent\textbf{Integrating Nudges into the Process.} To ensure participants' conformance with the to-be process, nudges can be included in their personal workflows. For example, participants can be guided to perform tasks differently once the to-be process is adopted. Nudges such as \emph{default} and \emph{framing} are relevant here.

\subsection{Implementation}

In the process implementation phase, the changes required to move from the as-is process to the to-be process are prepared and performed, covering both organizational change management and automation. There are threats in the (mis-)use of information systems in which processes are implemented. We describe two main opportunities when an organization decides to implement its processes and what type of nudging can be used for that.

\noindent\textbf{Addressing Potential Fears of Process Participants.} 
Implementing new processes and executing them through information systems often encounters resistance from stakeholders who fear the unknown. This makes it hard to properly gather the correct requirements needed for the implementation of the process. Some of the nudges that may tackle this resistance are \emph{default}, \emph{framing}, and \emph{simplification}.
\textit{Default} can be used to first describe the optimal process and then ask if the participants agree with this interpretation. \textit{Framing} can be used to better gather information from participants. For example, framing questions in a way that emphasizes the positive outcomes of the new process can shift the focus from resistance to cooperation. Instead of asking, "What issues might arise from this new process?" a framing approach would be, "How will this new process help us achieve our goals and improve sustainability?". Complementing this, \textit{simplification} may be used by providing participants with typical scenarios where they have to make a decision.

\noindent\textbf{Avoiding Misuse of New Information System.}
The introduction of new information systems carries the risk that employees might misuse or underuse them due to a lack of understanding or intentional manipulation. Nudging strategies can help mitigate these risks. For instance, \emph{salience} can guide the use of the system correctly. Visual cues such as color coding, bold text, or strategic placement can address the attention to critical features and usage guidelines. Additionally, \textit{default options} to make the steps of the new process the primary option can reduce resistance to change and reduce mistakes.  These default settings can also be used to support sustainable practices. For example, software systems can be pre-configured to optimize resource usage and reduce waste.

\subsection{Monitoring}
Process monitoring is the BPM phase in which data, typically stored in information systems, is used to analyze the status of the running business process. 

\noindent\textbf{Providing Transparent Feedback and Continuous Improvement.}
Providing transparent, real-time \emph{feedback} on performance, including sustainability metrics, can nudge continuous improvement. Dashboards that visualize environmental performance can motivate ongoing adherence to sustainable practices.

\noindent\textbf{Providing Timely Interventions Based on Monitoring Data.}
Using nudges such as timely \textit{reminders} (i.e., alerts and notifications) when performance deviates from sustainability goals can prompt quick corrective actions. This ensures that the processes remain aligned with both operational and environmental objectives.

\noindent\textbf{Supporting Data Collection and Reporting.} Collecting data and reporting includes extracting and manipulating data from information systems event logs and presenting the analysis of the results to the user. Data collection may suffer from unsystematic and non-diligent behavior of the analyst in charge. Additionally, reporting may be of low-quality when results do not explain actual issues or fail to point to actionable items. The following nudges may be used. \textit{Reminders} may help the analyst to act. This may be paired with defined data collection methodologies, such as~\cite{DBLP:conf/caise/EckLLA15}. For reporting, process mining results can be used to \emph{frame} adherence to conformant behavior~\cite{DBLP:conf/icpm/HobeckPW22}. To aid this, a \emph{feedback} nudge can be used, by providing direct information on the consequences of the the issues pointed out by process mining outputs.

\section{Implications of Nudging for Green BPM}
\label{sec:discussion}
Next, we turn to discuss the opportunities (O) and challenges (C) of nudging for Green BPM:

    \mypar{O1 - Better sustainability of processes} Nudging can be integrated into all phases of the BPM lifecycle.
    This allows us to incorporate ideas of sustainability within business processes. Therefore, nudging can be one tool towards sustainable business processes. 
    
    \mypar{O2 - Designing interventions for more sustainability} Nudges can not only lead to more sustainable business processes at design time, but can also be integrated at runtime; leading to a more sustainable behavior of process workers and customers. 
    Previous work showed how nudges can affect decisions for a desired outcome, and so far this type of intervention received limited attention in BPM.

    \mypar{C1 - Nudges do not guarantee a desired outcome} The effectiveness of nudges can vary significantly depending on the context and individual differences~\cite{hummel2019effective}. Research has shown that not all nudges have the same impact across different cultures ~\cite{johnson2003defaults}. Therefore, what works in one company might not work in another. In this scenario, the design and testing of nudges are essential to ensure they achieve the intended result.

    \mypar{C2 - Evaluating effectiveness} Evaluating whether a nudge is achieving the desired outcomes, such as increased sustainability, presents another challenge. The methods used to measure the effectiveness of the nudge should take into consideration short and long-term sustainability results. Continuous monitoring and adjustment important to maintain the nudge's effectiveness.

    \mypar{C3 - Not a sole strategy} Nudges should be seen as part of a broader strategy for promoting sustainable business processes. Relaying uniquely on nudging may jeopardize sustainability efforts. Other measures such as integrated sustainability management, eco-efficiency, sustainable innovation, and competitive advantage are crucial to have a holistic approach~\cite{danciu2013sustainable}.
    
    \mypar{C4 - Ethical considerations} One of the primary challenges in the application of nudging is the ethical implications of influencing behavior without explicit consent~\cite{sunstein2015ethics}. Thaler and Sunstein~\cite{thaler2021nudge} argue that nudges should be transparent and not manipulative, but in practice, the boundary between acceptable influence and manipulation can be blurry.  In the context of Green BPM, where the goal is to promote environmentally sustainable behavior, it is crucial to ensure that nudges do not infringe on individual autonomy or influence stakeholders to make unethical decisions. Therefore, transparency and preservation of freedom of choice must be an integral part of the implementation of the nudge. Additionally, nudges with illicit ends must be avoided even when there is general consent~\cite{sunstein2015ethics}.

\section{Conclusion} \label{sec:conclusion} 
In this paper, we showed how nudging can be used alongside the BPM lifecycle. Especially, we argue how nudges can be used to as a tool to achieve more sustainable business processes and process mining. 
We discuss that, while providing opportunities for inducing stakeholders towards desired positive behavior, nudging comes along with challenges and ethical considerations.
In future work, we plan to study specific instances of how nudges can be integrated into business process management tasks or specific business processes. 

\mypar{Acknowledgments} This research was supported by the Einstein Foundation Berlin under grant  EPP-2019-524 and the BMBF under grant 16DII133.

\bibliographystyle{splncs}
\bibliography{references.bib}
\end{document}